\documentclass[aps,prb,twocolumn,groupedaddress]{revtex4}
\usepackage{graphicx}
\usepackage{dcolumn}
\usepackage{bm}
\usepackage{color}
\begin{document}
%
%
\title{Fe-doping induced evolution of charge-orbital ordering in a bicritical-state manganite}
%

\author{H. Sakai$^1$, K. Ito$^1$, R. Kumai$^2$ and Y. Tokura$^{1, 2, 3}$}
\affiliation{$^1$Department of Applied Physics, University of Tokyo, Tokyo 113-8656, Japan\\
$^2$ Correlated Electron Research Center (CERC), National Institute of Advanced Industrial Science and Technology (AIST), Tsukuba 305-8562, Japan\\
$^3$Multiferroics Project, ERATO, Japan Science and Technology Agency (JST), Tokyo 113-8656, Japan}


%
\begin{abstract}
Impurity effects on the stability of a ferromagnetic metallic state in a bicritical-state manganite, (La$_{0.7}$Pr$_{0.3}$)$_{0.65}$Ca$_{0.35}$MnO$_3$, on the verge of metal-insulator transition have been investigated by substituting a variety of transition-metal atoms for Mn ones.
Among them, Fe doping exhibits the exceptional ability to dramatically decrease the ferromagnetic transition temperature.
Systematic studies on the magnetotransport properties and x-ray diffraction for the Fe-doped crystals have revealed that charge-orbital ordering evolves down to low temperatures, which strongly suppresses the ferromagnetic metallic state.
The observed glassy magnetic and transport properties as well as diffuse phase transition can be attributed to the phase-separated state where short-range charge-orbital-ordered clusters are embedded in the ferromagnetic metallic matrix.
Such a behavior in the Fe-doped manganites form a marked contrast to the Cr-doping effects on charge-orbital-ordered manganites known as impurity-induced collapse of charge-orbital ordering.
\end{abstract}
%
\date{\today}
%
\maketitle
%
\section{INTRODUCTION}
In hole-doped perovskite-type manganites, such as $R_{1-x}A_{x}$MnO$_3$ ($R$ being a trivalent rare-earth ion and $A$ being a divalent alkaline-earth ion), a ferromagnetic metallic (FM) phase and a charge-orbital-ordered antiferromagnetic insulating (CO/OOI) phase compete with each other.
Complex states of matter and critical changes among them, such as colossal magnetoresistance and magnetic-field-induced insulator-metal transition\cite{Tokura2006RPPa, Dagotto2001PRa}, manifest themselves in a bicritical region, where such competing phases are almost degenerate in free energy and separated by a first-order phase transition line.
Quenched disorder arising from local lattice distortions and/or dopant ions dramatically modifies the electronic structure near the bicritical point\cite{Dagotto2005Sciencea, Burgy2001PRLa, Burgy2004PRLa, Murakami2003PRLa, Motome2003PRLa, Pradhan2007PRLa}.
One typical example of moderate disorder is the solid solution of the perovskite A-site with $R$ and $A$ ions (of small and large ionic radii, respectively), which is a well-known and widely-used technique to control the band filling (hole doping level) and the one-electron bandwidth of the system\cite{Imada1998RMPa}.
Randomly distributed $R$ and $A$ ions result in the local lattice distortions and the random Coulomb potential, which can cause the suppression of long-range ordered state of both FM and CO/OOI\cite{Rodriguez1996PRBa, Rodriguez2000PRBa}.
Consequently, a {\it homogeneous} spin-glass insulating phase dominates in the bicritical region when the mismatch of ionic radii between $R$ and $A$ ions is appreciable\cite{Tomioka2003PRBa, Tomioka2004PRBa, Mathieu2004PRLa}.
%
\par
%
As another typical example of strong disorder, effects of impurity doping onto Mn sites have been intensively studied so far.
In particular, an impact of Cr doping in half-doped charge-orbital-ordered (CO/OO) manganites has attracted much attention for past years\cite{Barnabe1997APLa, Raveau1997JSSCa, Katsufuji1999JPSJa, Kimura1999PRLa, Martinelli2006PRBa}.
Only a few percent substitution of Cr for Mn dramatically suppresses the long-range CO/OO state and makes the system (partly) ferromagnetic metallic.
Since Cr$^{3+}$ is quite stable in valence, it may serve as the immovable $e_{g}$ charge-orbital deficiency in the CO/OO state and locally induce the competing FM state as the alternative, which leads to {\it inhomogeneous} phase separation with both CO/OOI and FM clusters coexisting on various length scales\cite{Mori2003PRBa}.
Other transition metals can also induce metallicity and ferromagnetism partially or completely in CO/OO manganites, depending on the kind of doping species\cite{Hebert2002SSComa, Machida2002PRBa, Yaicle2004SSComa, Maignan2001JAPb}.
Ru, Ni or Co doping, for example, easily suppresses the pristine CO/OO state and effectively stabilizes the FM state, as in the case of Cr doping.
For Fe, Ga or Al doping, on the other hand, no clear FM transition shows up in the absence of a magnetic field, but only a spin-glass-like (insulating) state is formed.
It was suggested that the electronic configuration of impurity ions may cause such a difference in the behavior of destroying the CO/OO state\cite{Hebert2002SSComa,Machida2002PRBa}.
%
\par
%
In this study, from the contrastive point of view, we have investigated effects of the disorder on the FM state by doping various transition-metal elements onto Mn sites.
Although impurity effects on many kinds of FM manganites were studied previously\cite{Ghosh1999PRBa, Ahn1996PRBa, Rubinstein1997PRBa, Blasco1997PRBa, Sun1999PRBa, Sun2000PRBa}, we focused here on the FM state of single-crystal (La$_{0.7}$Pr$_{0.3}$)$_{0.65}$Ca$_{0.35}$MnO$_3$, which locates near the phase boundary to the CO/OOI, to reveal the disorder effects on the FM phase near the bicritical point as complementary to those on the CO/OO phase.
The adopted compound (La$_{0.7}$Pr$_{0.3}$)$_{0.65}$Ca$_{0.35}$MnO$_3$ is close to the bicriticality but least affected by the disorder effect arising from the A-site solid solution, and hence can provide the ideal arena to highlight the genuine B-site doping effect on the FM state.
The stability of the FM state against impurity doping has been observed to strongly depend on the kind of dopants, as in the case of the CO/OO state.
Among all the dopants, Fe dopants most effectively decrease the FM transition temperature, $T_{\rm C}$ (by $\sim$70\% for 5\% doping).
In the Fe-doped manganites, we have observed the evolution of short-range charge-orbital ordering down to low temperatures, which strongly suppresses the FM state.
Such a tendency, which was revealed in this purposely designed compound but should be generic for the FM state in the bicritical-state manganites, markedly contrasts with the Cr-doping effects on the CO/OO manganites.
%
\section{EXPERIMENT}
Single crystals of (La$_{1-x}$Pr$_x$)$_{0.65}$Ca$_{0.35}$MnO$_3$ ($0.2\!\le\!x\!\le\!0.8$) and (La$_{0.7}$Pr$_{0.3}$)$_{0.65}$Ca$_{0.35}$Mn$_{1-y}${\it M}$_y$O$_3$ ({\it M}=Fe, Cr, Ga and Ru, $0\!\le\!y\!\le\!0.1$) were grown by the floating zone method.
Mixed powders of La$_2$O$_3$, Pr$_6$O$_{11}$, CaCO$_3$, Mn$_3$O$_4$, $\alpha$-Fe$_2$O$_3$, Cr$_2$O$_3$, Ga$_2$O$_3$ and RuO$_2$ in stoichiometric proportions were first calcined at 1000--1050$^\circ$C for 10-20 hours in air.
The mixture was pulverized and again sintered at 1200--1250$^\circ$C for 30--40 hours in air.
The resulting powders were pulverized and then pressed into a rod with $\sim$5 mm in diameter and $\sim$60 mm in length.
The rod was fired at 1350--1400$^\circ$C for 30--40 hours in air.
The crystal growths except for {\it M}=Ru were performed in an oxygen atmosphere with rotating the feed and seed rods in opposite directions at the rate of 15--20 rpm while the growth for {\it M}=Ru in air.
The growth rate was set at 2--2.5 mm/h.
Electron probe microanalysis (EPMA) revealed that the beginning part of the grown crystal rod has the composition variation along the growth direction, probably due to the chemical instability of the molten zone in an early process of the crystal growth.
The middle and end parts, on the other hand, have a homogeneous composition, from which we have prepared the samples for all the measurements performed in this study.
Furthermore, inductively coupled plasma (ICP) spectroscopy on the obtained samples has shown that their composition is equal to the prescribed ratio with an accuracy of $\pm0.007$ and $\pm0.003$ for $x$ and $y$, respectively.
For a Ru-doped compound, however, we had only a 2\%-Ru-doped crystal of (La$_{0.7}$Pr$_{0.3}$)$_{0.65}$Ca$_{0.35}$Mn$_{0.98}$Ru$_{0.02}$O$_3$ due to the high volatility of Ru oxides although we prescribed the ratio for 5\% doping of Ru in the mixed powders.
The powder x-ray diffraction patterns showed that the obtained crystals are of single phase and that the crystal structure is orthorhombic at room temperature with $a_0\sim b_0\sim c_0/\sqrt{2}\sim \sqrt{2}a_p$, where $a_p$ is the lattice parameter of the pseudocubic lattice.
Because the orthorhombic distortion is small and the twin domains equally exist, we here employ cubic notation for simplicity.
For several crystals, the synchrotron single-crystal x-ray diffraction was performed, using an imaging plate system on the beam line BL-1A of the Photon Factory, High-Energy Accelerator Research Organization (KEK), Japan.
Magnetization was measured with a SQUID magnetometer.
Resistivity was measured by a conventional four-probe method with electrodes formed by heat-treatment-type silver paint.
%
\section{RESULTS AND DISCUSSION}
%
\subsection{Bicritical features in (La$_{1-x}$Pr$_{x}$)$_{0.65}$Ca$_{0.35}$MnO$_3$ crystals}
%
\begin{figure}
\includegraphics[width=8cm]{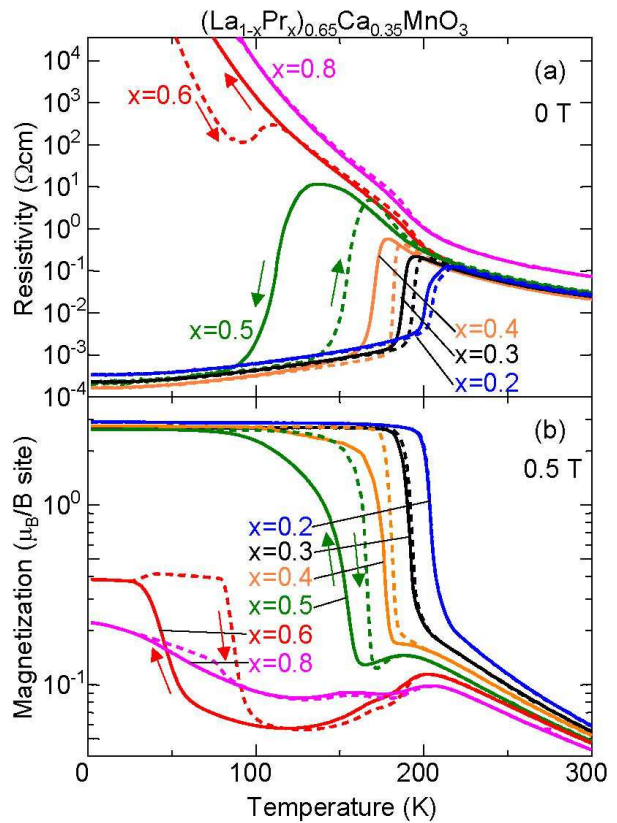}
\caption{\label{fig:LPCMOtemperature}(Color online) Temperature profiles of (a) resistivity at $\mu_0H\!=\!0$ T and (b) magnetization at $\mu_0H\!=\!0.5$ T for crystals of (La$_{1-x}$Pr$_{x}$)$_{0.65}$Ca$_{0.35}$MnO$_3$ ($0.2\!\le\!x\!\le\!0.8$). Solid and dashed lines indicate cooling and warming runs, respectively.}
\end{figure}
%
We first show the bicriticality in (La$_{1-x}$Pr$_{x}$)$_{0.65}$Ca$_{0.35}$MnO$_3$, which is the mixed crystal system of FM La$_{0.65}$Ca$_{0.35}$MnO$_3$ and CO/OOI Pr$_{0.65}$Ca$_{0.35}$MnO$_3$.
Figure \ref{fig:LPCMOtemperature} shows the temperature profiles of (a) resistivity at $\mu_0H\!=\!0$ T and (b) magnetization at $\mu_0H\!=\!0.5$ T for single crystals of (La$_{1-x}$Pr$_{x}$)$_{0.65}$Ca$_{0.35}$MnO$_3$ ($0.2\!\le\!x\!\le\!0.8$).
With increasing $x$, the FM transition accompanying a sharp drop of resistivity systematically decreases in $T_{\rm C}$ and the CO/OO state, which is manifested at around 200 K by a steep increase in resistivity or decrease in magnetization, shows up for $x\!\ge\!0.4$.
The re-entrant insulator (CO/OOI) -metal (FM) transition is observed with large temperature hysteresis for intermediate doping levels, $x\!=\!0.4$ and 0.5.
Finally, the CO/OOI phase dominates the region of $x\!\ge\!0.6$ and the charge-orbital ordering temperature, $T_{\rm CO}$, gradually increases as $x$ increases.
For $x\!=\!0.6$, the insulating low-temperature resistivity shows a small drop even at zero magnetic field, as is barely visible in a warming run.
This stems from that even tiny disorder or inhomogeneity in the sample could induce a small fraction of the FM state in the long-range CO/OOI phase since the $x\!=\!0.6$ crystal is just on their phase boundary at the ground state.
%
\par
\begin{figure}
\includegraphics[width=7.5cm]{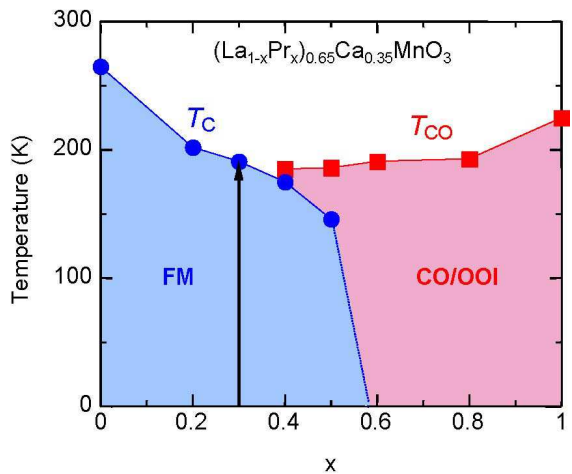}
\caption{\label{fig:LPCMOphase}(Color online) Phase diagram for (La$_{1-x}$Pr$_x$)$_{0.65}$Ca$_{0.35}$MnO$_3$. The ferromagnetic metal and charge-orbital-ordered insulator are denoted by FM and CO/OOI, respectively. $T_{\rm C}$ and $T_{\rm CO}$, which were determined in cooling runs, are denoted with closed circles and closed squares, respectively. An arrow shows the parent compound ($x\!=\!0.3$) for the impurity doping experiment described in this paper. The data for $x\!=\!0$ and 1 are from Ref. \onlinecite{Cheong1999booka} (poly crystal) and Ref. \onlinecite{Tomioka2002PRBa} (single crystal), respectively.}
\end{figure}

%
Figure \ref{fig:LPCMOphase} summarizes the electronic phase diagram for the (La$_{1-x}$Pr$_{x}$)$_{0.65}$Ca$_{0.35}$MnO$_3$ ($0\!\le\!x\!\le\!1$) crystals.
$T_{\rm C}$ is determined as the temperature where the resistivity curve in Fig. \ref{fig:LPCMOtemperature}(a) shows a steep drop, while $T_{\rm CO}$ is where $|d\ln\rho/dT|$ curve shows a cusp-like maximum.
For clarity, only the transition temperatures in a cooling process are shown in Fig. \ref{fig:LPCMOphase}.
The FM phase is replaced with the CO/OOI phase with increasing $x$.
Note that the FM region stretches out in the CO/OOI region below $\sim$150 K, reflecting the FM re-entrant transition as observed for $x\!=\!0.4$ and 0.5 in Fig. \ref{fig:LPCMOtemperature}.
The end-point of the FM phase is around $x\!=\!0.6$.
As a result, the phase diagram for the (La$_{1-x}$Pr$_{x}$)$_{0.65}$Ca$_{0.35}$MnO$_3$ system exhibits an almost ideal bicritical feature arising from the competition between the long-range CO/OOI and FM as previously reported for Pr$_{0.65}$(Ca$_{1-x}$Sr$_{x}$)$_{0.35}$MnO$_3$\cite{Tomioka1997JPSJa, Tomioka2002PRBa}.
Because the mismatch of the ionic size between trivalent rare-earth (La and Pr) and divalent alkaline-earth (Ca) is the smallest among a series of $R_{0.65}A_{0.35}$MnO$_3$\cite{Tomioka2004PRBa}, the quenched disorder arising from A-site solid solution can be kept minimal in magnitude to maintain either of the long-range orders in the vicinity of the bicritical point for this system.
%
\par
%
%
We can now adopt (La$_{1-x}$Pr$_{x}$)$_{0.65}$Ca$_{0.35}$MnO$_3$ as one of the most ideal and typical parent compounds for investigating the impurity doping effects on the bicritical regime.
Hereinafter, to show the effects on the FM states close to the bicritical point, we restrict ourselves to the case of $x\!=\!0.3$ (FM), as denoted with the vertical arrow in Fig. \ref{fig:LPCMOphase}.
We have partially substituted the Mn sites with various transition-metal elements as a strong quenched disorder in this system.
All the crystals investigated here are single crystals to exclude the effects from the extrinsic disorder such as grain boundaries and crystal lattice defects, which often dominates the properties in polycrystalline crystals\cite{Uehara1999Naturea}.
%
\subsection{Stability of ferromagnetic metallic state against a variety of impurity species}
%
\begin{figure}
\includegraphics[width=8cm]{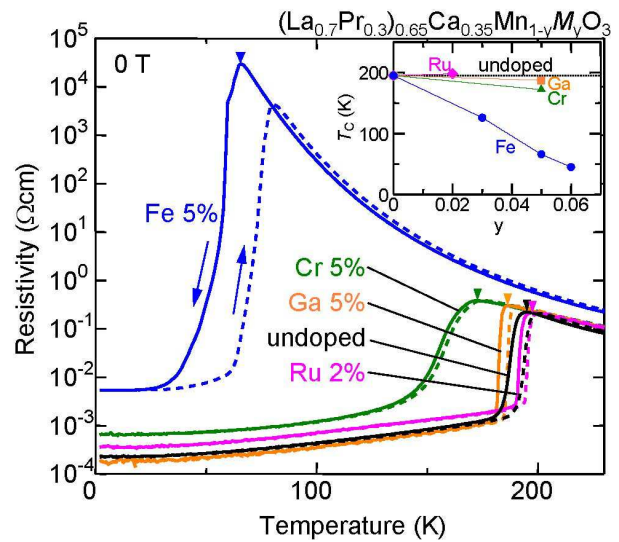}
\caption{\label{fig:dopant}(Color online) Temperature profiles of resistivity at $\mu_0H\!=\!0$ T for undoped, 5\% {\it M}-doped (La$_{0.7}$Pr$_{0.3}$)$_{0.65}$Ca$_{0.35}$Mn$_{0.95}$$M$$_{0.05}$O$_3$ ($M$=Fe, Cr and Ga) and 2\% Ru-doped (La$_{0.7}$Pr$_{0.3}$)$_{0.65}$Ca$_{0.35}$Mn$_{0.98}$Ru$_{0.02}$O$_3$ manganites. The peaks, corresponding to $T_{\rm C}$, are denoted with closed triangles. Inset: $T_{\rm C}$ versus the doping level for the above crystals.}
\end{figure}
%
Figure \ref{fig:dopant} shows the temperature dependence of the resistivity for doped FM manganites, (La$_{0.7}$Pr$_{0.3}$)$_{0.65}$Ca$_{0.35}$Mn$_{0.95}$$M$$_{0.05}$O$_3$ ($M$=Mn, Fe, Cr and Ga) and (La$_{0.7}$Pr$_{0.3}$)$_{0.65}$Ca$_{0.35}$Mn$_{0.98}$Ru$_{0.02}$O$_3$.
The pristine (La$_{0.7}$Pr$_{0.3}$)$_{0.65}$Ca$_{0.35}$MnO$_3$ is FM at the ground state, as seen by the presence of a steep drop at $T_{\rm C}\!\sim\!195$ K in the resistivity curve.
In the Fe-doped crystal, even for only 5\% doping, the FM state is strongly destabilized and $T_{\rm C}$ decreases down to $\sim$66 K, where the resistivity drops by $\sim$6 orders of magnitude.
For 5\% doping of Cr and Ga, however, impurity effects are much smaller and $T_{\rm C}$ is slightly lowered to $\sim$172 K and $\sim$187 K, respectively.
Note that the changes in the carrier density due to doping are almost the same among the above 5\% doped crystals since the trivalent states are stable for all the dopant species.
The electronic (spin) configurations, however, are different, such as $t_{2g}^{3}\,e_{g}^{2}$ ($S\!=\!5/2$), $t_{2g}^{3}\,e_{g}^{0}$ ($S\!=\!3/2$) and $t_{2g}^{6}\,e_{g}^{4}$ ($S\!=\!0$) for Fe$^{3+}$, Cr$^{3+}$ and Ga$^{3+}$, respectively, which could play an important role in destabilizing the FM state.
The Ru-doped crystal, on the other hand, exhibits even a slight increase in $T_{\rm C}$, as is often observed in other FM manganites\cite{Weigand2002APLa, Onose2005APLa, Yamada2005APLa}.
Inset in Fig. \ref{fig:dopant} summarizes shifts in $T_{\rm C}$ versus the doping level for the doped FM manganites.
%
\par
%
Impurity effects on the FM states thus markedly change from one dopant to the other, as in the case of the CO/OO states\cite{Hebert2002SSComa, Machida2002PRBa, Yaicle2004SSComa, Maignan2001JAPa}.
This reflects that the competing ordered phases in the bicritical-state manganites, FM and CO/OOI, are selectively favored or disfavored depending on the impurity species.
In fact, although the substitutions of Cr or Ru\cite{Maignan2001JAPa, Maignan2001JAPb, Raveau2000JSSCa, Martin2001PRBa} strongly suppress the CO/OO state while locally inducing the FM state, they have no significant influence on the FM state to give merely small changes in $T_{\rm C}$ as shown above.
Conversely, we here find the exceptional ability of Fe doping to suppress the FM state and drastically decrease $T_{\rm C}$.
Hereinafter, we concentrate on systematic studies on Fe-doping effects by measurements of resistivity, magnetization and x-ray diffraction.
%
\subsection{Transport and magnetic properties}
%
\begin{figure}
\includegraphics[width=8cm]{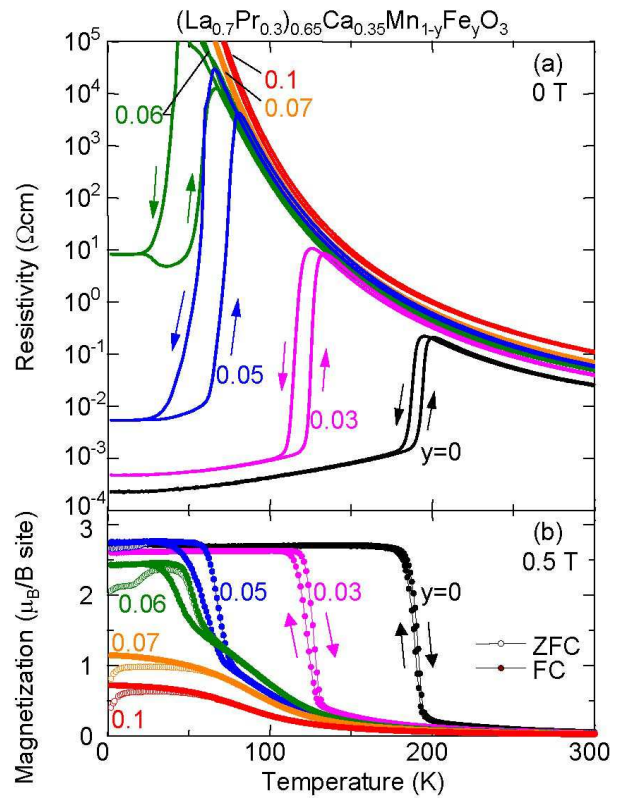}
\caption{\label{fig:LPCMFOtemperature}(Color online) Temperature profiles of (a) resistivity at $\mu_0H\!=\!0$ T and (b) magnetization at $\mu_0H\!=\!0.5$ T for crystals of (La$_{0.7}$Pr$_{0.3}$)$_{0.65}$Ca$_{0.35}$Mn$_{1-y}$Fe$_y$O$_3$ ($0\!\le\!y\!\le\!0.1$). Measurements were performed in both cooling and warming runs. The magnetization data measured after the field-cooling and zero-field-cooling processes are denoted with the closed circles (FC) and the open ones (ZFC), respectively.}
\end{figure}
%
We show in Fig. \ref{fig:LPCMFOtemperature} the temperature profiles of (a) resistivity and (b) magnetization for the crystals of (La$_{0.7}$Pr$_{0.3}$)$_{0.65}$Ca$_{0.35}$Mn$_{1-y}$Fe$_y$O$_3$ ($0\!\le\!y\!\le\!0.1$).
Measurements for the resistivity and magnetization were carried out at $\mu_{0}H\!=\!0$ T and 0.5 T, respectively, in both cooling and warming runs.
The magnetization data measured in a warming run after cooling the samples at $\mu_{0}H\!=\!0$ T are denoted with the open circles (ZFC) while those measured after cooling them at $\mu_{0}H\!=\!0.5$ T are with the close circles (FC).
A few percent substitutes of Fe effectively suppress the FM state and $T_{\rm C}$ systematically decreases with the increase of Fe-doping level.
The $y\!\le\!0.06$ crystals show the steep resistivity drop around $T_{\rm C}$ with temperature hysteresis and become metallic below $T_{\rm C}$.
The residual resistivity, $\rho_{0}$, however, increases by a factor of $\sim$10$^4$ with increasing $y$ from 0 to 0.06.
Note that the value of $\rho_{0}\!\sim\!8.3\ \Omega$cm for the $y\!=\!0.06$ crystal is too large, far above the Ioffe-Regel limit $\sim\!10^{-3}\ \Omega$cm, for homogeneous FM states.
Therefore, this suggests that the observed insulator-metal transition is attributed to the percolation transition in the inhomogeneous system where the FM clusters randomly exist.
In the crystals with $y\!\ge\!0.07$, the FM transition completely disappears and the resistivity shows insulating down to the lowest temperature.
The temperature profiles of magnetization for the $y\!\ge\!0.07$ crystals show the distinct history dependence between zero-field cooling and field cooling below $\sim$50 K, as shown in Fig. \ref{fig:LPCMFOtemperature}(b), which indicates that a spin-glass state becomes dominant at low temperatures for these crystals.
Even in the crystals with $y\!=\!0.05$ and 0.06, which show the FM transition, such a discrepancy between zero-field-cooled and field-cooled magnetization is also observed below $\sim$50 K.
This behavior reflects that the ground state for these crystals is not a homogeneous FM, but an inhomogeneous mixture composed of the FM and CO/OO states, as evidenced by the diffraction study ({\it vide infra}).
%
\subsection{Single-crystal x-ray diffraction}
%
\begin{figure}
\includegraphics[width=8cm]{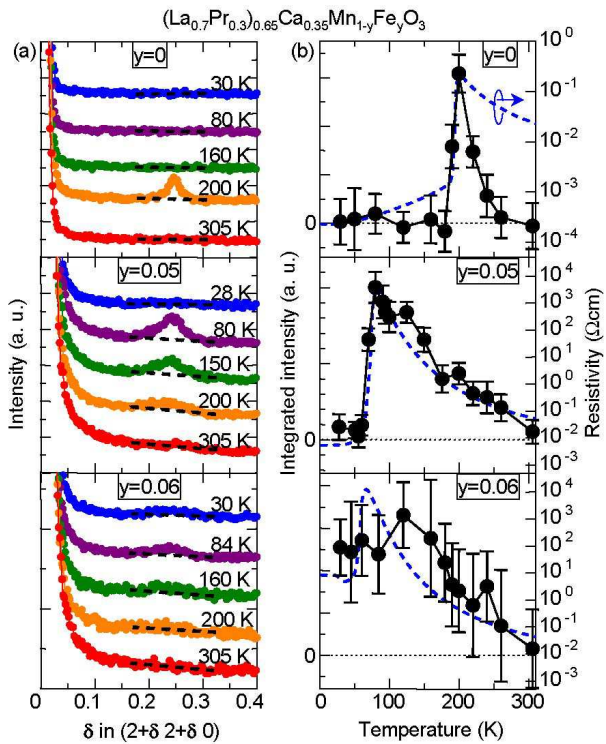}
\caption{\label{fig:xrayprofile}(Color online) (a) Temperature dependence of x-ray diffraction profiles along [110] direction around the (200) Bragg peak for (La$_{0.7}$Pr$_{0.3}$)$_{0.65}$Ca$_{0.35}$Mn$_{1-y}$Fe$_y$O$_3$ ($y\!=\!$0, 0.05 and 0.06) crystals. The closed circles represent the data points and the dashed lines indicate the estimated backgrounds including the intensity of the diffuse scattering. (b) Temperature profiles of the integrated intensity of the x-ray superlattice reflection related to the charge-orbital ordering around a wave vector of ($\frac{1}{4}\frac{1}{4}0$) for the $y\!=\!0$, $y\!=\!0.05$ and $y\!=\!0.06$ crystals. The dashed lines represent the corresponding temperature profiles of resistivity. These measurements were performed in a warming run.}
\end{figure}
%
The single-crystal x-ray diffraction measurements for the crystals with $y\!=\!0$, 0.05 and 0.06 have revealed the close relation between the strong suppression of the FM state and the evolution of the CO/OO state. 
Figure \ref{fig:xrayprofile}(a) shows the x-ray diffraction profiles along [110] direction around the (200) Bragg point at several temperatures, where we find a CO/OO-related superlattice reflection with a modulation vector of $\sim$($\frac{1}{4}\frac{1}{4}0$) in the cubic lattice setting.
We also display in Fig. \ref{fig:xrayprofile}(b) the temperature dependence of the integrated intensity of the superlattice reflection for each crystal.
For comparison, the corresponding resistivity curves are shown there with dashed lines.
In the undoped crystal with $y\!=\!0$, the CO/OO-related superlattice peak shows up only just above $T_{\rm C}$ and completely disappears below $T_{\rm C}$.
With the increase of $y$ from 0 to 0.05, the superlattice reflection intensity continues to increase to lower temperatures, but it suddenly decreases at $T_{\rm C}$, which is in accord with the resistive transition.
In the $y\!=\!0.06$ crystal, on the other hand, the superlattice peak is found to persist down to the lowest temperature while the resistivity shows a steep drop around $T_{\rm C}$, which suggests that both FM and CO/OOI phases coexist at the ground state.
Note that the correlation length of charge-orbital order as extracted from the superlattice peak width for $y\!=\!0.06$ is $\sim$20\AA, indicating that the CO/OO state survives with only short-range correlation, such as in a form of clusters embedded in the FM matrix.
%
\par
%
With the further increase of Fe-doping level ($y\!\ge\!0.07$), the clear CO/OO superlattice reflection gradually disappears with the broadened peak width, while the diffuse scattering (Huang scattering) extending in the direction of [110] and [1$\bar{1}$0] around the Bragg peak becomes dominant.
Such diffuse scattering arises from the less-correlated polarons associated with the Jahn-Teller distortion due to the localization of carriers on Mn sites\cite{Shimomura1999PRLa, Doloc1999PRLa, Shimomura2000PRBa}, which can be regarded as the remnant of the CO/OO superlattice spot\cite{Kimura2000PRBa}.
In fact, the temperature dependence of the intensity of the diffuse scattering is quite similar to that of the superlattice reflection in the crystals with $y\!=\!0.05$ and 0.06, in which the both are observed, as shown in Fig. \ref{fig:xrayprofile}(a) (middle and bottom panels).
In the crystals with $y\!\ge\!0.07$, we have found that the diffuse scattering (with a weak peak structure of the superlattice reflection) survives down to the lowest temperature.
This indicates that the CO/OO correlation robustly remains at the ground state even for $y\!\ge\!0.07$, although it is weakened to be extremely short range and/or dynamical because of the high concentration of Fe substitution causing the strong quenched disorder.
%
\subsection{Phase diagram for Fe-doped (La$_{0.7}$Pr$_{0.3}$)$_{0.65}$Ca$_{0.35}$MnO$_3$}
%
\begin{figure}
\includegraphics[width=8cm]{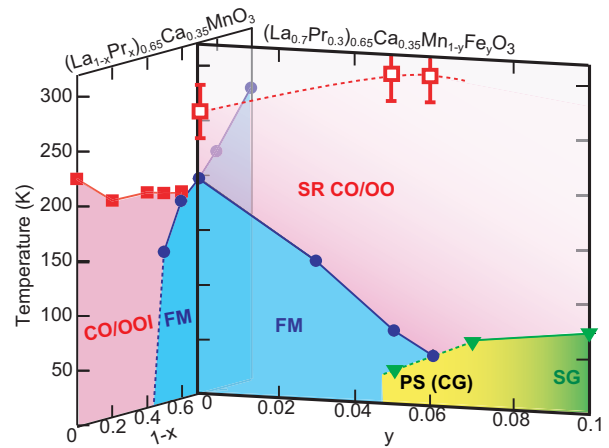}
\caption{\label{fig:LPCMFOphase}(Color online) Phase diagrams for (La$_{1-x}$Pr$_x$)$_{0.65}$Ca$_{0.35}$MnO$_3$ and (La$_{0.7}$Pr$_{0.3}$)$_{0.65}$Ca$_{0.35}$Mn$_{1-y}$Fe$_y$O$_3$. The phase-separated (PS) state, or cluster glass (CG), where the ferromagnetic-metallic (FM) clusters and the short-range charge-orbital-ordered (SR CO/OO) clusters coexist, appears around $y\!\sim\!0.06$. With further increasing $y$, it changes to the homogeneous spin-glass (SG) with atomic-scale correlation. The transition temperature to the FM, CO/OO and SG (CG) phase is denoted with closed circles, squares and triangles, respectively. The onset temperature of SR CO/OO, determined by the integrated intensity of x-ray superlattice reflection, is denoted with open squares.}
\end{figure}
%
Figure \ref{fig:LPCMFOphase} shows the electronic phase diagram for (La$_{0.7}$Pr$_{0.3}$)$_{0.65}$Ca$_{0.35}$Mn$_{1-y}$Fe$_y$O$_3$ ($0\!\le\! y\!\le\!0.1$) combined with that for (La$_{1-x}$Pr$_{x}$)$_{0.65}$Ca$_{0.35}$MnO$_3$ ($0\!\le\! x\!\le\!1$).
The transition temperature to the spin glass (SG) is determined as the temperature below which the magnetization shows a history dependence between field cooling and zero-field cooling.
The onset temperature of short-range (SR) CO/OO is estimated from Fig. \ref{fig:xrayprofile}.
In the pristine crystal of (La$_{0.7}$Pr$_{0.3}$)$_{0.65}$Ca$_{0.35}$MnO$_3$, which locates near the phase boundary to the CO/OOI, the ground state is a homogeneous FM.
The SR CO/OO manifests itself only in a narrow temperature range just above $T_{\rm C}$ and suddenly vanishes below $T_{\rm C}$.
As $y$ increases, the FM state is suppressed with a steep decrease in $T_{\rm C}$ and the SR CO/OO instead evolves down to lower temperatures.
The onset temperature of SR CO/OO shows no significant change regardless of $y$, and hence may be interpreted as a Griffiths temperature\cite{Burgy2001PRLa, Dagotto2005NJPa}.
At around $y\!\sim\!0.06$, the clear FM transition disappears and the ground state becomes an inhomogeneously phase-separated one, where the FM and CO/OO clusters coexist even at the lowest temperature, which can often be regarded as cluster glass (CG).
With further increasing $y$, however, the correlation length of the CO/OO state as well as the FM state becomes shorter due to the strong randomness from highly concentrated Fe doping.
Near $y\!\sim\!0.1$, finally, the homogeneous spin-glass insulating phase with only atomic-scale correlation prevails below $\sim$50 K.
%
\par
%
As shown above, Fe doping has a large impact on the FM state; it effectively weakens the FM state and instead develops the CO/OO state down to low temperatures.
In the CO/OO state, however, Fe substitution seems to play only minor roles\cite{Hebert2002SSComa, Machida2002PRBa}.
Although a small fraction of the FM state could be induced at low temperatures for a tiny doping of Fe ($\sim$2--4\%) in some cases\cite{Damay1997JAPa,Levy2001JMMMa}, it is easily extinguished with further doping (up to as low as $\sim$4--7\%).
These suggest that Fe substitution selectively hinders the FM correlation rather than the CO/OO one.
Such a tendency contrasts strikingly with that in the Cr-doping effects on the bicritical-state manganites.
A few percent substitutes of Cr in the CO/OO manganite strongly destroy the long-range CO/OO and induce the FM state, as investigated intensively\cite{Barnabe1997APLa, Raveau1997JSSCa, Katsufuji1999JPSJa, Kimura1999PRLa}.
Our study reveals the contrastive feature that Cr doping has only small effects on the FM state while slightly decreasing $T_{\rm C}$ (see Fig. \ref{fig:dopant}).
Since both Fe and Cr ions prefer to be in trivalent states in the perovskite oxides, effects of the carrier density change cannot explain the contrastive impacts between Fe and Cr doping.
In addition, the both ions are coupled antiferromagnetically with Mn ones\cite{Ahn1996PRBa, Ogale1998PRBa, Simopoulos1999PRBa, Studer1999JJAPa}.
The microscopic origin for the difference should be seeked for in the electronic configurations of doped ions, i. e., Fe$^{3+}$ ($t_{2g}^{3}\,e_{g}^{2}$) and Cr$^{3+}$ ($t_{2g}^{3}\,e_{g}^{0}$), which may give rise to the different nature of the disorder effect.
Fe$^{3+}$ with the occupied $e_{g}$ state may be apt to localize $e_{g}$ electrons of Mn$^{3+}$ via strong antiferromagnetic superexchange interaction between them.
Meanwhile, Cr$^{3+}$ with the empty $e_{g}$ state does not strongly interfere with the conduction of $e_{g}$ electrons, which can instead have a large effect on the CO/OO state by the $t_{2g}$ superexchange interaction, which will favorably break the antiferromagnetic coupling between the FM chains in the CE-type structure\cite{Martin2001PRBa}.
In other words, the disorder effect from the doped Fe$^{3+}$ ions mainly acts on the charge sector of the Mn conduction electrons, while that from the doped Cr$^{3+}$ ions on the spin sector.
The quantitative account for the microscopic origin of the marked contrasts between Fe and Cr doping, however, remains to be explored as a future problem.
%
\subsection{Relaxor ferromagnet behavior in the $y\!=\!0.07$ crystal}
%
\begin{figure}
\includegraphics[width=8cm]{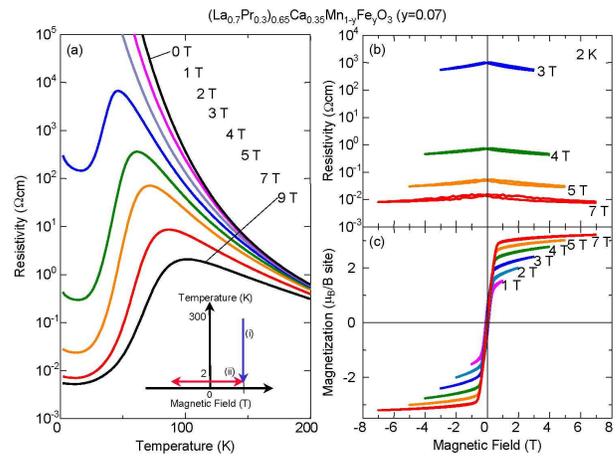}
\caption{\label{fig:relaxor}(Color online) (a) Temperature profiles of resistivity at various magnetic fields for the (La$_{0.7}$Pr$_{0.3}$)$_{0.65}$Ca$_{0.35}$Mn$_{1-y}$Fe$_y$O$_3$ ($y\!=\!0.07$) crystal. Magnetic-field dependence of (b) resistivity and (c) magnetization at 2 K after the magnetic-field annealing procedures for the $y\!=\!0.07$ crystal. The measurements were performed once the magnetic fields were applied at room temperature and the crystal was cooled down to 2 K at the respective annealing fields, as depicted in the inset.}
\end{figure}
%
In the following, we focus on the $y\!=\!0.07$ crystal with the phase-separated ground state in which the FM conducting matrix is barely disconnected by the short-range CO/OO clusters (see Figs. \ref{fig:LPCMFOtemperature} and \ref{fig:LPCMFOphase}), that is, the volume fraction of the FM state will be just below the percolation threshold.
We show in Fig. \ref{fig:relaxor}(a) the temperature profiles of resistivity for the $y\!=\!0.07$ crystal at various external magnetic fields.
At zero field, the resistivity shows an insulating behavior down to the lowest temperature.
The application of a magnetic field of a few tesla, however, drastically reduces the resistivity at low temperatures, making the system metallic.
More noteworthy is that the value of the residual resistivity varies in a very wide range depending on the magnitude of the magnetic field.
We have further investigated such a multistable behavior by measuring the resistivity and magnetization at the lowest temperature of 2 K by changing the strength of the magnetic field, as shown in Figs. \ref{fig:relaxor}(b) and \ref{fig:relaxor}(c), respectively.
For this measurement with the procedure of ``magnetic-field annealing"\cite{Kimura1999PRLa, Kimura2001JAP}, we first apply various magnetic fields $H$ at room temperature and then cool the sample down to 2 K.
Subsequently, we sweep the magnetic field between $+H$ and $-H$, as displayed in the inset of Fig. \ref{fig:relaxor}(a).
The resistivity dramatically decreases by more than five orders of magnitude with increasing $\mu_{0}H$ from 0 to 9 T, although those at $\mu_{0}H\!\le\!2$ T are too high to measure.
Once after a temperature is lowered down to 2 K at $H$, the values of the resistivity are little changed (on such a logarithmic scale) even after the magnetic field is removed, memorizing the history of the annealing process.
In the corresponding magnetization curves, furthermore, the saturation moment steadily increases with the increase of $H$.
This systematic change in the resistivity and magnetization can be attributed to the change in the volume fraction of the CO/OO clusters embedded in the FM matrix.
A magnetic-field annealing at $\mu_{0}H\!\ge\!7$ T sufficiently suppresses the CO/OO clusters at the ground state, making the system almost homogeneous FM.
Thus, we can easily control the volume fraction of the CO/OO clusters in the phase-separated state by changing the magnitude of $H$.
Analogous glassy magnetotransport properties or {\it relaxor ferromagnet} phenomena have been reported for Cr-doped CO/OO manganites\cite{Kimura1999PRLa, Kimura2001JAP}, where the FM phase is locally induced in the CO/OO matrix conversely to the present case.
Thus, the phase-separated states, whose respective phase volume can be critically controlled by magnetic fields, are generated from the two very contrastive ordered states, FM and CO/OOI, in the bicritical-state manganites.
%
\section{CONCLUSION}
%
In conclusion, we have investigated the impurity doping effects on a ferromagnetic metallic state in single crystals of (La$_{0.7}$Pr$_{0.3}$)$_{0.65}$Ca$_{0.35}$MnO$_3$, which locate near the bicritical point between the long-range charge-orbital-ordered state and ferromagnetic state.
The effect of doping on the ferromagnetic metallic state strongly depends on the electronic configurations of the doped impurities, such as Fe, Cr, Ga and Ru.
For example, 5\% doping of Cr or Ga leads to a merely small decrease in $T_{\rm C}$ ($-\Delta T_{C}/T_{C}\!<\!0.12$), while Ru doping rather stabilizes the ferromagnetic state ($+\Delta T_{C}/T_{C}\!\sim\!0.02$ for 2\% doping).
The most noteworthy is that a few percent doping of Fe significantly decreases $T_{\rm C}$.
By measuring magnetotransport properties and x-ray diffraction, we have revealed that Fe doping induces the (short-range) charge-orbital ordering down to low temperatures and strongly suppresses the ferromagnetic metallic state.
Such a behavior contrasts strikingly with the Cr-doping effects on the charge-orbital-ordered manganites.
Furthermore, we have observed relaxor-like (glassy) magnetotransport properties, such as magnetic-field annealing effects, which reflects the phase-separated inhomogeneous state composed of the charge-orbital-ordered clusters embedded in the ferromagnetic metallic matrix, in the Fe-doped (e.g. 7\%) crystal.
Thus, the present systematic study on the impurity substitution effects demonstrates a variety of methods for the phase control in the bicritical-state manganites with impurity doping as well as external (magnetic) fields.
%
\begin{acknowledgments}
%
We thank S. Miyasaka and Y. Onose for fruitful discussion.
We also thank Y. Kiuchi for her help in ICP measurements performed in the Materials Design and Characterization Laboratory, Institute for Solid State Physics, University of Tokyo.
The support to H. S. by the 21st Century COE Program for ``Applied Physics on Strong Correlation'' is appreciated.
This study was partly supported by JSPS KAKENHI (No. 15104006) and MEXT TOKUTEI (No. 16076205).
%
\end{acknowledgments}
\bibliography{fdico_resub2}

\begin{thebibliography}{51}
\expandafter\ifx\csname natexlab\endcsname\relax\def\natexlab#1{#1}\fi
\expandafter\ifx\csname bibnamefont\endcsname\relax
  \def\bibnamefont#1{#1}\fi
\expandafter\ifx\csname bibfnamefont\endcsname\relax
  \def\bibfnamefont#1{#1}\fi
\expandafter\ifx\csname citenamefont\endcsname\relax
  \def\citenamefont#1{#1}\fi
\expandafter\ifx\csname url\endcsname\relax
  \def\url#1{\texttt{#1}}\fi
\expandafter\ifx\csname urlprefix\endcsname\relax\def\urlprefix{URL }\fi
\providecommand{\bibinfo}[2]{#2}
\providecommand{\eprint}[2][]{\url{#2}}

\bibitem[{\citenamefont{Tokura}(2006)}]{Tokura2006RPPa}
\bibinfo{author}{\bibfnamefont{Y.}~\bibnamefont{Tokura}},
  \bibinfo{journal}{Rep.\ Prog.\ Phys.} \textbf{\bibinfo{volume}{69}},
  \bibinfo{pages}{797} (\bibinfo{year}{2006}).

\bibitem[{\citenamefont{Dagotto et~al.}(2001)\citenamefont{Dagotto, Hotta, and
  Moreo}}]{Dagotto2001PRa}
\bibinfo{author}{\bibfnamefont{E.}~\bibnamefont{Dagotto}},
  \bibinfo{author}{\bibfnamefont{T.}~\bibnamefont{Hotta}}, \bibnamefont{and}
  \bibinfo{author}{\bibfnamefont{A.}~\bibnamefont{Moreo}},
  \bibinfo{journal}{Phys.\ Rep.} \textbf{\bibinfo{volume}{344}},
  \bibinfo{pages}{1} (\bibinfo{year}{2001}).

\bibitem[{\citenamefont{Dagotto}(2005{\natexlab{a}})}]{Dagotto2005Sciencea}
\bibinfo{author}{\bibfnamefont{E.}~\bibnamefont{Dagotto}},
  \bibinfo{journal}{Science} \textbf{\bibinfo{volume}{309}},
  \bibinfo{pages}{257} (\bibinfo{year}{2005}{\natexlab{a}}).

\bibitem[{\citenamefont{Burgy et~al.}(2001)\citenamefont{Burgy, Mayr,
  Martin-Mayor, Moreo, and Dagotto}}]{Burgy2001PRLa}
\bibinfo{author}{\bibfnamefont{J.}~\bibnamefont{Burgy}},
  \bibinfo{author}{\bibfnamefont{M.}~\bibnamefont{Mayr}},
  \bibinfo{author}{\bibfnamefont{V.}~\bibnamefont{Martin-Mayor}},
  \bibinfo{author}{\bibfnamefont{A.}~\bibnamefont{Moreo}}, \bibnamefont{and}
  \bibinfo{author}{\bibfnamefont{E.}~\bibnamefont{Dagotto}},
  \bibinfo{journal}{Phys.\ Rev.\ Lett.} \textbf{\bibinfo{volume}{87}},
  \bibinfo{pages}{277202} (\bibinfo{year}{2001}).

\bibitem[{\citenamefont{Burgy et~al.}(2004)\citenamefont{Burgy, Moreo, and
  Dagotto}}]{Burgy2004PRLa}
\bibinfo{author}{\bibfnamefont{J.}~\bibnamefont{Burgy}},
  \bibinfo{author}{\bibfnamefont{A.}~\bibnamefont{Moreo}}, \bibnamefont{and}
  \bibinfo{author}{\bibfnamefont{E.}~\bibnamefont{Dagotto}},
  \bibinfo{journal}{Phys.\ Rev.\ Lett.} \textbf{\bibinfo{volume}{92}},
  \bibinfo{pages}{97202} (\bibinfo{year}{2004}).

\bibitem[{\citenamefont{Murakami and Nagaosa}(2003)}]{Murakami2003PRLa}
\bibinfo{author}{\bibfnamefont{S.}~\bibnamefont{Murakami}} \bibnamefont{and}
  \bibinfo{author}{\bibfnamefont{N.}~\bibnamefont{Nagaosa}},
  \bibinfo{journal}{Phys.\ Rev.\ Lett.} \textbf{\bibinfo{volume}{90}},
  \bibinfo{pages}{197201} (\bibinfo{year}{2003}).

\bibitem[{\citenamefont{Motome et~al.}(2003)\citenamefont{Motome, Furukawa, and
  Nagaosa}}]{Motome2003PRLa}
\bibinfo{author}{\bibfnamefont{Y.}~\bibnamefont{Motome}},
  \bibinfo{author}{\bibfnamefont{N.}~\bibnamefont{Furukawa}}, \bibnamefont{and}
  \bibinfo{author}{\bibfnamefont{N.}~\bibnamefont{Nagaosa}},
  \bibinfo{journal}{Phys.\ Rev.\ Lett.} \textbf{\bibinfo{volume}{91}},
  \bibinfo{pages}{167204} (\bibinfo{year}{2003}).

\bibitem[{\citenamefont{Pradhan et~al.}(2007)\citenamefont{Pradhan, Mukherjee,
  and Majumdar}}]{Pradhan2007PRLa}
\bibinfo{author}{\bibfnamefont{K.}~\bibnamefont{Pradhan}},
  \bibinfo{author}{\bibfnamefont{A.}~\bibnamefont{Mukherjee}},
  \bibnamefont{and} \bibinfo{author}{\bibfnamefont{P.}~\bibnamefont{Majumdar}},
  \bibinfo{journal}{Phys.\ Rev.\ Lett.} \textbf{\bibinfo{volume}{99}},
  \bibinfo{pages}{147206} (\bibinfo{year}{2007}).

\bibitem[{\citenamefont{Imada et~al.}(1998)\citenamefont{Imada, Fujimori, and
  Tokura}}]{Imada1998RMPa}
\bibinfo{author}{\bibfnamefont{M.}~\bibnamefont{Imada}},
  \bibinfo{author}{\bibfnamefont{A.}~\bibnamefont{Fujimori}}, \bibnamefont{and}
  \bibinfo{author}{\bibfnamefont{Y.}~\bibnamefont{Tokura}},
  \bibinfo{journal}{Rev.\ Mod.\ Phys.} \textbf{\bibinfo{volume}{70}},
  \bibinfo{pages}{1039} (\bibinfo{year}{1998}).

\bibitem[{\citenamefont{Rodriguez-Martinez and
  Attfield}(1996)}]{Rodriguez1996PRBa}
\bibinfo{author}{\bibfnamefont{L.~M.} \bibnamefont{Rodriguez-Martinez}}
  \bibnamefont{and} \bibinfo{author}{\bibfnamefont{J.~P.}
  \bibnamefont{Attfield}}, \bibinfo{journal}{Phys.\ Rev.\ B}
  \textbf{\bibinfo{volume}{54}}, \bibinfo{pages}{R15622}
  (\bibinfo{year}{1996}).

\bibitem[{\citenamefont{Rodriguez-Martinez and
  Attfield}(2000)}]{Rodriguez2000PRBa}
\bibinfo{author}{\bibfnamefont{L.~M.} \bibnamefont{Rodriguez-Martinez}}
  \bibnamefont{and} \bibinfo{author}{\bibfnamefont{J.~P.}
  \bibnamefont{Attfield}}, \bibinfo{journal}{Phys.\ Rev.\ B}
  \textbf{\bibinfo{volume}{63}}, \bibinfo{pages}{024424}
  (\bibinfo{year}{2000}).

\bibitem[{\citenamefont{Tomioka et~al.}(2003)\citenamefont{Tomioka, Okimoto,
  Jung, Kumai, and Tokura}}]{Tomioka2003PRBa}
\bibinfo{author}{\bibfnamefont{Y.}~\bibnamefont{Tomioka}},
  \bibinfo{author}{\bibfnamefont{Y.}~\bibnamefont{Okimoto}},
  \bibinfo{author}{\bibfnamefont{J.~H.} \bibnamefont{Jung}},
  \bibinfo{author}{\bibfnamefont{R.}~\bibnamefont{Kumai}}, \bibnamefont{and}
  \bibinfo{author}{\bibfnamefont{Y.}~\bibnamefont{Tokura}},
  \bibinfo{journal}{Phys.\ Rev.\ B} \textbf{\bibinfo{volume}{68}},
  \bibinfo{pages}{094417} (\bibinfo{year}{2003}).

\bibitem[{\citenamefont{Tomioka and Tokura}(2004)}]{Tomioka2004PRBa}
\bibinfo{author}{\bibfnamefont{Y.}~\bibnamefont{Tomioka}} \bibnamefont{and}
  \bibinfo{author}{\bibfnamefont{Y.}~\bibnamefont{Tokura}},
  \bibinfo{journal}{Phys.\ Rev.\ B} \textbf{\bibinfo{volume}{70}},
  \bibinfo{pages}{014432} (\bibinfo{year}{2004}).

\bibitem[{\citenamefont{Mathieu et~al.}(2004)\citenamefont{Mathieu, Akahoshi,
  Asamitsu, Tomioka, and Tokura}}]{Mathieu2004PRLa}
\bibinfo{author}{\bibfnamefont{R.}~\bibnamefont{Mathieu}},
  \bibinfo{author}{\bibfnamefont{D.}~\bibnamefont{Akahoshi}},
  \bibinfo{author}{\bibfnamefont{A.}~\bibnamefont{Asamitsu}},
  \bibinfo{author}{\bibfnamefont{Y.}~\bibnamefont{Tomioka}}, \bibnamefont{and}
  \bibinfo{author}{\bibfnamefont{Y.}~\bibnamefont{Tokura}},
  \bibinfo{journal}{Phys.\ Rev.\ Lett.} \textbf{\bibinfo{volume}{93}},
  \bibinfo{pages}{227202} (\bibinfo{year}{2004}).

\bibitem[{\citenamefont{Bernabe et~al.}(1997)\citenamefont{Bernabe, Maignan,
  Hervieu, Damay, Martin, and Raveau}}]{Barnabe1997APLa}
\bibinfo{author}{\bibfnamefont{A.}~\bibnamefont{Bernabe}},
  \bibinfo{author}{\bibfnamefont{A.}~\bibnamefont{Maignan}},
  \bibinfo{author}{\bibfnamefont{M.}~\bibnamefont{Hervieu}},
  \bibinfo{author}{\bibfnamefont{F.}~\bibnamefont{Damay}},
  \bibinfo{author}{\bibfnamefont{C.}~\bibnamefont{Martin}}, \bibnamefont{and}
  \bibinfo{author}{\bibfnamefont{B.}~\bibnamefont{Raveau}},
  \bibinfo{journal}{Appl.\ Phys.\ Lett.} \textbf{\bibinfo{volume}{71}},
  \bibinfo{pages}{3907} (\bibinfo{year}{1997}).

\bibitem[{\citenamefont{Raveau et~al.}(1997)\citenamefont{Raveau, Maignan, and
  Martin}}]{Raveau1997JSSCa}
\bibinfo{author}{\bibfnamefont{B.}~\bibnamefont{Raveau}},
  \bibinfo{author}{\bibfnamefont{A.}~\bibnamefont{Maignan}}, \bibnamefont{and}
  \bibinfo{author}{\bibfnamefont{C.}~\bibnamefont{Martin}},
  \bibinfo{journal}{J.\ Solid\ State\ Chem.} \textbf{\bibinfo{volume}{130}},
  \bibinfo{pages}{162} (\bibinfo{year}{1997}).

\bibitem[{\citenamefont{Katsufuji et~al.}(1999)\citenamefont{Katsufuji, Cheong,
  Mori, and Chen}}]{Katsufuji1999JPSJa}
\bibinfo{author}{\bibfnamefont{T.}~\bibnamefont{Katsufuji}},
  \bibinfo{author}{\bibfnamefont{S.-W.} \bibnamefont{Cheong}},
  \bibinfo{author}{\bibfnamefont{S.}~\bibnamefont{Mori}}, \bibnamefont{and}
  \bibinfo{author}{\bibfnamefont{C.~H.} \bibnamefont{Chen}},
  \bibinfo{journal}{J.\ Phys.\ Soc.\ Jpn.} \textbf{\bibinfo{volume}{68}},
  \bibinfo{pages}{1090} (\bibinfo{year}{1999}).

\bibitem[{\citenamefont{Kimura et~al.}(1999)\citenamefont{Kimura, Tomioka,
  Kumai, Okimoto, and Tokura}}]{Kimura1999PRLa}
\bibinfo{author}{\bibfnamefont{T.}~\bibnamefont{Kimura}},
  \bibinfo{author}{\bibfnamefont{Y.}~\bibnamefont{Tomioka}},
  \bibinfo{author}{\bibfnamefont{R.}~\bibnamefont{Kumai}},
  \bibinfo{author}{\bibfnamefont{Y.}~\bibnamefont{Okimoto}}, \bibnamefont{and}
  \bibinfo{author}{\bibfnamefont{Y.}~\bibnamefont{Tokura}},
  \bibinfo{journal}{Phys.\ Rev.\ Lett.} \textbf{\bibinfo{volume}{83}},
  \bibinfo{pages}{3940} (\bibinfo{year}{1999}).

\bibitem[{\citenamefont{Martinelli et~al.}(2006)\citenamefont{Martinelli,
  Ferretti, Castellano, Mondelli, Cimberle, Tropeano, and
  Ritter}}]{Martinelli2006PRBa}
\bibinfo{author}{\bibfnamefont{A.}~\bibnamefont{Martinelli}},
  \bibinfo{author}{\bibfnamefont{M.}~\bibnamefont{Ferretti}},
  \bibinfo{author}{\bibfnamefont{C.}~\bibnamefont{Castellano}},
  \bibinfo{author}{\bibfnamefont{C.}~\bibnamefont{Mondelli}},
  \bibinfo{author}{\bibfnamefont{M.~R.} \bibnamefont{Cimberle}},
  \bibinfo{author}{\bibfnamefont{M.}~\bibnamefont{Tropeano}}, \bibnamefont{and}
  \bibinfo{author}{\bibfnamefont{C.}~\bibnamefont{Ritter}},
  \bibinfo{journal}{Phys.\ Rev.\ B} \textbf{\bibinfo{volume}{73}},
  \bibinfo{pages}{064423} (\bibinfo{year}{2006}).

\bibitem[{\citenamefont{Mori et~al.}(2003)\citenamefont{Mori, Shoji, Yamamoto,
  Asaka, Matsui, Machida, Moritomo, and Katsufuji}}]{Mori2003PRBa}
\bibinfo{author}{\bibfnamefont{S.}~\bibnamefont{Mori}},
  \bibinfo{author}{\bibfnamefont{R.}~\bibnamefont{Shoji}},
  \bibinfo{author}{\bibfnamefont{N.}~\bibnamefont{Yamamoto}},
  \bibinfo{author}{\bibfnamefont{T.}~\bibnamefont{Asaka}},
  \bibinfo{author}{\bibfnamefont{Y.}~\bibnamefont{Matsui}},
  \bibinfo{author}{\bibfnamefont{A.}~\bibnamefont{Machida}},
  \bibinfo{author}{\bibfnamefont{Y.}~\bibnamefont{Moritomo}}, \bibnamefont{and}
  \bibinfo{author}{\bibfnamefont{T.}~\bibnamefont{Katsufuji}},
  \bibinfo{journal}{Phys.\ Rev.\ B} \textbf{\bibinfo{volume}{67}},
  \bibinfo{pages}{012403} (\bibinfo{year}{2003}).

\bibitem[{\citenamefont{H$\rm\acute{e}$bert
  et~al.}(2002)\citenamefont{H$\rm\acute{e}$bert, Maignan, Martin, and
  Raveau}}]{Hebert2002SSComa}
\bibinfo{author}{\bibfnamefont{S.}~\bibnamefont{H$\rm\acute{e}$bert}},
  \bibinfo{author}{\bibfnamefont{A.}~\bibnamefont{Maignan}},
  \bibinfo{author}{\bibfnamefont{C.}~\bibnamefont{Martin}}, \bibnamefont{and}
  \bibinfo{author}{\bibfnamefont{B.}~\bibnamefont{Raveau}},
  \bibinfo{journal}{Solid State Commun.} \textbf{\bibinfo{volume}{121}},
  \bibinfo{pages}{229} (\bibinfo{year}{2002}).

\bibitem[{\citenamefont{Machida et~al.}(2002)\citenamefont{Machida, Moritomo,
  Ohoyama, Katsufuji, and Nakamura}}]{Machida2002PRBa}
\bibinfo{author}{\bibfnamefont{A.}~\bibnamefont{Machida}},
  \bibinfo{author}{\bibfnamefont{Y.}~\bibnamefont{Moritomo}},
  \bibinfo{author}{\bibfnamefont{K.}~\bibnamefont{Ohoyama}},
  \bibinfo{author}{\bibfnamefont{T.}~\bibnamefont{Katsufuji}},
  \bibnamefont{and} \bibinfo{author}{\bibfnamefont{A.}~\bibnamefont{Nakamura}},
  \bibinfo{journal}{Phys.\ Rev.\ B} \textbf{\bibinfo{volume}{65}},
  \bibinfo{pages}{64435} (\bibinfo{year}{2002}).

\bibitem[{\citenamefont{Yaicle et~al.}(2004)\citenamefont{Yaicle, Raveau,
  Maignan, and Hervieu}}]{Yaicle2004SSComa}
\bibinfo{author}{\bibfnamefont{C.}~\bibnamefont{Yaicle}},
  \bibinfo{author}{\bibfnamefont{B.}~\bibnamefont{Raveau}},
  \bibinfo{author}{\bibfnamefont{A.}~\bibnamefont{Maignan}}, \bibnamefont{and}
  \bibinfo{author}{\bibfnamefont{M.}~\bibnamefont{Hervieu}},
  \bibinfo{journal}{Solid State Commun.} \textbf{\bibinfo{volume}{132}},
  \bibinfo{pages}{487} (\bibinfo{year}{2004}).

\bibitem[{\citenamefont{Maignan
  et~al.}(2001{\natexlab{a}})\citenamefont{Maignan, Martin, Hervieu, Raveau,
  and Hejtmanek}}]{Maignan2001JAPb}
\bibinfo{author}{\bibfnamefont{A.}~\bibnamefont{Maignan}},
  \bibinfo{author}{\bibfnamefont{C.}~\bibnamefont{Martin}},
  \bibinfo{author}{\bibfnamefont{M.}~\bibnamefont{Hervieu}},
  \bibinfo{author}{\bibfnamefont{B.}~\bibnamefont{Raveau}}, \bibnamefont{and}
  \bibinfo{author}{\bibfnamefont{J.}~\bibnamefont{Hejtmanek}},
  \bibinfo{journal}{J.\ Appl.\ Phys.} \textbf{\bibinfo{volume}{89}},
  \bibinfo{pages}{2232} (\bibinfo{year}{2001}{\natexlab{a}}).

\bibitem[{\citenamefont{Ghosh et~al.}(1999)\citenamefont{Ghosh, Ogale, Ramesh,
  Greene, Venkatesan, Gapchup, Bathe, and Patil}}]{Ghosh1999PRBa}
\bibinfo{author}{\bibfnamefont{K.}~\bibnamefont{Ghosh}},
  \bibinfo{author}{\bibfnamefont{S.~B.} \bibnamefont{Ogale}},
  \bibinfo{author}{\bibfnamefont{R.}~\bibnamefont{Ramesh}},
  \bibinfo{author}{\bibfnamefont{R.~L.} \bibnamefont{Greene}},
  \bibinfo{author}{\bibfnamefont{T.}~\bibnamefont{Venkatesan}},
  \bibinfo{author}{\bibfnamefont{K.~M.} \bibnamefont{Gapchup}},
  \bibinfo{author}{\bibfnamefont{R.}~\bibnamefont{Bathe}}, \bibnamefont{and}
  \bibinfo{author}{\bibfnamefont{S.~I.} \bibnamefont{Patil}},
  \bibinfo{journal}{Phys.\ Rev.\ B} \textbf{\bibinfo{volume}{59}},
  \bibinfo{pages}{533} (\bibinfo{year}{1999}).

\bibitem[{\citenamefont{Ahn et~al.}(1996)\citenamefont{Ahn, Wu, Liu, and
  Chien}}]{Ahn1996PRBa}
\bibinfo{author}{\bibfnamefont{K.~H.} \bibnamefont{Ahn}},
  \bibinfo{author}{\bibfnamefont{X.~W.} \bibnamefont{Wu}},
  \bibinfo{author}{\bibfnamefont{K.}~\bibnamefont{Liu}}, \bibnamefont{and}
  \bibinfo{author}{\bibfnamefont{C.~L.} \bibnamefont{Chien}},
  \bibinfo{journal}{Phys.\ Rev.\ B} \textbf{\bibinfo{volume}{54}},
  \bibinfo{pages}{15299} (\bibinfo{year}{1996}).

\bibitem[{\citenamefont{Rubinstein et~al.}(1997)\citenamefont{Rubinstein,
  Gillespie, Snyder, and Tritt}}]{Rubinstein1997PRBa}
\bibinfo{author}{\bibfnamefont{M.}~\bibnamefont{Rubinstein}},
  \bibinfo{author}{\bibfnamefont{D.~J.} \bibnamefont{Gillespie}},
  \bibinfo{author}{\bibfnamefont{J.~E.} \bibnamefont{Snyder}},
  \bibnamefont{and} \bibinfo{author}{\bibfnamefont{T.~M.} \bibnamefont{Tritt}},
  \bibinfo{journal}{Phys.\ Rev.\ B} \textbf{\bibinfo{volume}{56}},
  \bibinfo{pages}{5412} (\bibinfo{year}{1997}).

\bibitem[{\citenamefont{Blasco et~al.}(1997)\citenamefont{Blasco,
  Garc$\rm\acute{i}$a, de~Teresa, Ibarra, Perez, Algarabel, Marquina, and
  Ritter}}]{Blasco1997PRBa}
\bibinfo{author}{\bibfnamefont{J.}~\bibnamefont{Blasco}},
  \bibinfo{author}{\bibfnamefont{J.}~\bibnamefont{Garc$\rm\acute{i}$a}},
  \bibinfo{author}{\bibfnamefont{J.~M.} \bibnamefont{de~Teresa}},
  \bibinfo{author}{\bibfnamefont{M.~R.} \bibnamefont{Ibarra}},
  \bibinfo{author}{\bibfnamefont{J.}~\bibnamefont{Perez}},
  \bibinfo{author}{\bibfnamefont{P.~A.} \bibnamefont{Algarabel}},
  \bibinfo{author}{\bibfnamefont{C.}~\bibnamefont{Marquina}}, \bibnamefont{and}
  \bibinfo{author}{\bibfnamefont{C.}~\bibnamefont{Ritter}},
  \bibinfo{journal}{Phys.\ Rev.\ B} \textbf{\bibinfo{volume}{55}},
  \bibinfo{pages}{8905} (\bibinfo{year}{1997}).

\bibitem[{\citenamefont{Sun et~al.}(1999)\citenamefont{Sun, Xu, Zheng, and
  Zhang}}]{Sun1999PRBa}
\bibinfo{author}{\bibfnamefont{Y.}~\bibnamefont{Sun}},
  \bibinfo{author}{\bibfnamefont{X.}~\bibnamefont{Xu}},
  \bibinfo{author}{\bibfnamefont{L.}~\bibnamefont{Zheng}}, \bibnamefont{and}
  \bibinfo{author}{\bibfnamefont{Y.}~\bibnamefont{Zhang}},
  \bibinfo{journal}{Phys.\ Rev.\ B} \textbf{\bibinfo{volume}{60}},
  \bibinfo{pages}{12317} (\bibinfo{year}{1999}).

\bibitem[{\citenamefont{Sun et~al.}(2000)\citenamefont{Sun, Xu, and
  Zhang}}]{Sun2000PRBa}
\bibinfo{author}{\bibfnamefont{Y.}~\bibnamefont{Sun}},
  \bibinfo{author}{\bibfnamefont{X.}~\bibnamefont{Xu}}, \bibnamefont{and}
  \bibinfo{author}{\bibfnamefont{Y.}~\bibnamefont{Zhang}},
  \bibinfo{journal}{Phys.\ Rev.\ B} \textbf{\bibinfo{volume}{63}},
  \bibinfo{pages}{54404} (\bibinfo{year}{2000}).

\bibitem[{\citenamefont{Tomioka et~al.}(1997)\citenamefont{Tomioka, Asamitsu,
  Kuwahara, and Tokura}}]{Tomioka1997JPSJa}
\bibinfo{author}{\bibfnamefont{Y.}~\bibnamefont{Tomioka}},
  \bibinfo{author}{\bibfnamefont{A.}~\bibnamefont{Asamitsu}},
  \bibinfo{author}{\bibfnamefont{H.}~\bibnamefont{Kuwahara}}, \bibnamefont{and}
  \bibinfo{author}{\bibfnamefont{Y.}~\bibnamefont{Tokura}},
  \bibinfo{journal}{J.\ Phys.\ Soc.\ Jpn.} \textbf{\bibinfo{volume}{66}},
  \bibinfo{pages}{302} (\bibinfo{year}{1997}).

\bibitem[{\citenamefont{Tomioka and Tokura}(2002)}]{Tomioka2002PRBa}
\bibinfo{author}{\bibfnamefont{Y.}~\bibnamefont{Tomioka}} \bibnamefont{and}
  \bibinfo{author}{\bibfnamefont{Y.}~\bibnamefont{Tokura}},
  \bibinfo{journal}{Phys.\ Rev.\ B} \textbf{\bibinfo{volume}{66}},
  \bibinfo{pages}{104416} (\bibinfo{year}{2002}).

\bibitem[{\citenamefont{Cheong and Hwang}(1999)}]{Cheong1999booka}
\bibinfo{author}{\bibfnamefont{S.-W.} \bibnamefont{Cheong}} \bibnamefont{and}
  \bibinfo{author}{\bibfnamefont{H.~Y.} \bibnamefont{Hwang}}, in
  \emph{\bibinfo{booktitle}{Colossal Magnetoresistance Oxides}}, edited by
  \bibinfo{editor}{\bibfnamefont{Y.}~\bibnamefont{Tokura}}
  (\bibinfo{publisher}{Gordon and Breach, London}, \bibinfo{year}{1999}).

\bibitem[{\citenamefont{Uehara et~al.}(1999)\citenamefont{Uehara, Mori, Chen,
  and Cheong}}]{Uehara1999Naturea}
\bibinfo{author}{\bibfnamefont{M.}~\bibnamefont{Uehara}},
  \bibinfo{author}{\bibfnamefont{S.}~\bibnamefont{Mori}},
  \bibinfo{author}{\bibfnamefont{C.~H.} \bibnamefont{Chen}}, \bibnamefont{and}
  \bibinfo{author}{\bibfnamefont{S.-W.} \bibnamefont{Cheong}},
  \bibinfo{journal}{Nature} \textbf{\bibinfo{volume}{399}},
  \bibinfo{pages}{560} (\bibinfo{year}{1999}).

\bibitem[{\citenamefont{Weigand et~al.}(2002)\citenamefont{Weigand, Gold,
  Schmid, Geissler, Goering, Dorr, Krabbes, and Ruck}}]{Weigand2002APLa}
\bibinfo{author}{\bibfnamefont{F.}~\bibnamefont{Weigand}},
  \bibinfo{author}{\bibfnamefont{S.}~\bibnamefont{Gold}},
  \bibinfo{author}{\bibfnamefont{A.}~\bibnamefont{Schmid}},
  \bibinfo{author}{\bibfnamefont{J.}~\bibnamefont{Geissler}},
  \bibinfo{author}{\bibnamefont{Goering}},
  \bibinfo{author}{\bibfnamefont{K.}~\bibnamefont{Dorr}},
  \bibinfo{author}{\bibfnamefont{G.}~\bibnamefont{Krabbes}}, \bibnamefont{and}
  \bibinfo{author}{\bibfnamefont{K.}~\bibnamefont{Ruck}},
  \bibinfo{journal}{Appl.\ Phys.\ Lett.} \textbf{\bibinfo{volume}{81}},
  \bibinfo{pages}{2035} (\bibinfo{year}{2002}).

\bibitem[{\citenamefont{Onose et~al.}(2005)\citenamefont{Onose, He, Kaneko,
  Arima, and Tokura}}]{Onose2005APLa}
\bibinfo{author}{\bibfnamefont{Y.}~\bibnamefont{Onose}},
  \bibinfo{author}{\bibfnamefont{J.~P.} \bibnamefont{He}},
  \bibinfo{author}{\bibfnamefont{Y.}~\bibnamefont{Kaneko}},
  \bibinfo{author}{\bibfnamefont{T.}~\bibnamefont{Arima}}, \bibnamefont{and}
  \bibinfo{author}{\bibfnamefont{Y.}~\bibnamefont{Tokura}},
  \bibinfo{journal}{Appl.\ Phys.\ Lett.} \textbf{\bibinfo{volume}{86}},
  \bibinfo{pages}{242502} (\bibinfo{year}{2005}).

\bibitem[{\citenamefont{Yamada et~al.}(2005)\citenamefont{Yamada, Kawasaki, and
  Tokura}}]{Yamada2005APLa}
\bibinfo{author}{\bibfnamefont{H.}~\bibnamefont{Yamada}},
  \bibinfo{author}{\bibfnamefont{M.}~\bibnamefont{Kawasaki}}, \bibnamefont{and}
  \bibinfo{author}{\bibfnamefont{Y.}~\bibnamefont{Tokura}},
  \bibinfo{journal}{Appl.\ Phys.\ Lett.} \textbf{\bibinfo{volume}{86}},
  \bibinfo{pages}{192505} (\bibinfo{year}{2005}).

\bibitem[{\citenamefont{Maignan
  et~al.}(2001{\natexlab{b}})\citenamefont{Maignan, Martin, Hervieu, and
  Raveau}}]{Maignan2001JAPa}
\bibinfo{author}{\bibfnamefont{A.}~\bibnamefont{Maignan}},
  \bibinfo{author}{\bibfnamefont{C.}~\bibnamefont{Martin}},
  \bibinfo{author}{\bibfnamefont{M.}~\bibnamefont{Hervieu}}, \bibnamefont{and}
  \bibinfo{author}{\bibfnamefont{B.}~\bibnamefont{Raveau}},
  \bibinfo{journal}{J.\ Appl.\ Phys.} \textbf{\bibinfo{volume}{89}},
  \bibinfo{pages}{500} (\bibinfo{year}{2001}{\natexlab{b}}).

\bibitem[{\citenamefont{Raveau et~al.}(2000)\citenamefont{Raveau, Maignan,
  Martin, Mahendiran, and Hervieu}}]{Raveau2000JSSCa}
\bibinfo{author}{\bibfnamefont{B.}~\bibnamefont{Raveau}},
  \bibinfo{author}{\bibfnamefont{A.}~\bibnamefont{Maignan}},
  \bibinfo{author}{\bibfnamefont{C.}~\bibnamefont{Martin}},
  \bibinfo{author}{\bibfnamefont{R.}~\bibnamefont{Mahendiran}},
  \bibnamefont{and} \bibinfo{author}{\bibfnamefont{M.}~\bibnamefont{Hervieu}},
  \bibinfo{journal}{J.\ Solid\ State\ Chem.} \textbf{\bibinfo{volume}{151}},
  \bibinfo{pages}{330} (\bibinfo{year}{2000}).

\bibitem[{\citenamefont{Martin et~al.}(2001)\citenamefont{Martin, Maignan,
  Hervieu, Autret, Raveau, and Khomskii}}]{Martin2001PRBa}
\bibinfo{author}{\bibfnamefont{C.}~\bibnamefont{Martin}},
  \bibinfo{author}{\bibfnamefont{A.}~\bibnamefont{Maignan}},
  \bibinfo{author}{\bibfnamefont{M.}~\bibnamefont{Hervieu}},
  \bibinfo{author}{\bibfnamefont{C.}~\bibnamefont{Autret}},
  \bibinfo{author}{\bibfnamefont{B.}~\bibnamefont{Raveau}}, \bibnamefont{and}
  \bibinfo{author}{\bibfnamefont{D.~I.} \bibnamefont{Khomskii}},
  \bibinfo{journal}{Phys.\ Rev.\ B} \textbf{\bibinfo{volume}{63}},
  \bibinfo{pages}{174402} (\bibinfo{year}{2001}).

\bibitem[{\citenamefont{Shimomura et~al.}(1999)\citenamefont{Shimomura,
  Wakabayashi, Kuwahara, and Tokura}}]{Shimomura1999PRLa}
\bibinfo{author}{\bibfnamefont{S.}~\bibnamefont{Shimomura}},
  \bibinfo{author}{\bibfnamefont{N.}~\bibnamefont{Wakabayashi}},
  \bibinfo{author}{\bibfnamefont{H.}~\bibnamefont{Kuwahara}}, \bibnamefont{and}
  \bibinfo{author}{\bibfnamefont{Y.}~\bibnamefont{Tokura}},
  \bibinfo{journal}{Phys.\ Rev.\ Lett.} \textbf{\bibinfo{volume}{83}},
  \bibinfo{pages}{4389} (\bibinfo{year}{1999}).

\bibitem[{\citenamefont{Vasiliu-Doloc et~al.}(1999)\citenamefont{Vasiliu-Doloc,
  Rosenkranz, Osborn, Sinha, Lynn, Mesot, Seeck, Preosti, Fedro, and
  Mitchell}}]{Doloc1999PRLa}
\bibinfo{author}{\bibfnamefont{L.}~\bibnamefont{Vasiliu-Doloc}},
  \bibinfo{author}{\bibfnamefont{S.}~\bibnamefont{Rosenkranz}},
  \bibinfo{author}{\bibfnamefont{R.}~\bibnamefont{Osborn}},
  \bibinfo{author}{\bibfnamefont{S.~K.} \bibnamefont{Sinha}},
  \bibinfo{author}{\bibfnamefont{J.~W.} \bibnamefont{Lynn}},
  \bibinfo{author}{\bibfnamefont{J.}~\bibnamefont{Mesot}},
  \bibinfo{author}{\bibfnamefont{O.~H.} \bibnamefont{Seeck}},
  \bibinfo{author}{\bibfnamefont{G.}~\bibnamefont{Preosti}},
  \bibinfo{author}{\bibfnamefont{A.~J.} \bibnamefont{Fedro}}, \bibnamefont{and}
  \bibinfo{author}{\bibfnamefont{J.~F.} \bibnamefont{Mitchell}},
  \bibinfo{journal}{Phys.\ Rev.\ Lett.} \textbf{\bibinfo{volume}{83}},
  \bibinfo{pages}{4393} (\bibinfo{year}{1999}).

\bibitem[{\citenamefont{Shimomura et~al.}(2000)\citenamefont{Shimomura,
  Tonegawa, Tajima, Wakabayashi, Ikeda, Shobu, Noda, Tomioka, and
  Tokura}}]{Shimomura2000PRBa}
\bibinfo{author}{\bibfnamefont{S.}~\bibnamefont{Shimomura}},
  \bibinfo{author}{\bibfnamefont{T.}~\bibnamefont{Tonegawa}},
  \bibinfo{author}{\bibfnamefont{K.}~\bibnamefont{Tajima}},
  \bibinfo{author}{\bibfnamefont{N.}~\bibnamefont{Wakabayashi}},
  \bibinfo{author}{\bibfnamefont{N.}~\bibnamefont{Ikeda}},
  \bibinfo{author}{\bibfnamefont{T.}~\bibnamefont{Shobu}},
  \bibinfo{author}{\bibfnamefont{Y.}~\bibnamefont{Noda}},
  \bibinfo{author}{\bibfnamefont{Y.}~\bibnamefont{Tomioka}}, \bibnamefont{and}
  \bibinfo{author}{\bibfnamefont{Y.}~\bibnamefont{Tokura}},
  \bibinfo{journal}{Phys.\ Rev.\ B} \textbf{\bibinfo{volume}{62}},
  \bibinfo{pages}{3875} (\bibinfo{year}{2000}).

\bibitem[{\citenamefont{Kimura et~al.}(2000)\citenamefont{Kimura, Kumai,
  Okimoto, Tomioka, and Tokura}}]{Kimura2000PRBa}
\bibinfo{author}{\bibfnamefont{T.}~\bibnamefont{Kimura}},
  \bibinfo{author}{\bibfnamefont{R.}~\bibnamefont{Kumai}},
  \bibinfo{author}{\bibfnamefont{Y.}~\bibnamefont{Okimoto}},
  \bibinfo{author}{\bibfnamefont{Y.}~\bibnamefont{Tomioka}}, \bibnamefont{and}
  \bibinfo{author}{\bibfnamefont{Y.}~\bibnamefont{Tokura}},
  \bibinfo{journal}{Phys.\ Rev.\ B} \textbf{\bibinfo{volume}{62}},
  \bibinfo{pages}{15021} (\bibinfo{year}{2000}).

\bibitem[{\citenamefont{Dagotto}(2005{\natexlab{b}})}]{Dagotto2005NJPa}
\bibinfo{author}{\bibfnamefont{E.}~\bibnamefont{Dagotto}},
  \bibinfo{journal}{New J.\ Phys.} \textbf{\bibinfo{volume}{7}},
  \bibinfo{pages}{67} (\bibinfo{year}{2005}{\natexlab{b}}).

\bibitem[{\citenamefont{Damay et~al.}(1997)\citenamefont{Damay, Maignan,
  Martin, and Raveau}}]{Damay1997JAPa}
\bibinfo{author}{\bibfnamefont{F.}~\bibnamefont{Damay}},
  \bibinfo{author}{\bibfnamefont{A.}~\bibnamefont{Maignan}},
  \bibinfo{author}{\bibfnamefont{C.}~\bibnamefont{Martin}}, \bibnamefont{and}
  \bibinfo{author}{\bibfnamefont{B.}~\bibnamefont{Raveau}},
  \bibinfo{journal}{J.\ Appl.\ Phys.} \textbf{\bibinfo{volume}{82}},
  \bibinfo{pages}{1485} (\bibinfo{year}{1997}).

\bibitem[{\citenamefont{Levy et~al.}(2001)\citenamefont{Levy, Granja,
  Indelicato, Vega, Polla, and Parisi}}]{Levy2001JMMMa}
\bibinfo{author}{\bibfnamefont{P.}~\bibnamefont{Levy}},
  \bibinfo{author}{\bibfnamefont{L.}~\bibnamefont{Granja}},
  \bibinfo{author}{\bibfnamefont{E.}~\bibnamefont{Indelicato}},
  \bibinfo{author}{\bibfnamefont{D.}~\bibnamefont{Vega}},
  \bibinfo{author}{\bibfnamefont{G.}~\bibnamefont{Polla}}, \bibnamefont{and}
  \bibinfo{author}{\bibfnamefont{F.}~\bibnamefont{Parisi}},
  \bibinfo{journal}{J.\ Magn.\ Magn.\ Mater.} \textbf{\bibinfo{volume}{226}},
  \bibinfo{pages}{794} (\bibinfo{year}{2001}).

\bibitem[{\citenamefont{Ogale et~al.}(1998)\citenamefont{Ogale, Shreekala,
  Bathe, Date, Patil, Hannoyer, Petit, and Marest}}]{Ogale1998PRBa}
\bibinfo{author}{\bibfnamefont{S.~B.} \bibnamefont{Ogale}},
  \bibinfo{author}{\bibfnamefont{R.}~\bibnamefont{Shreekala}},
  \bibinfo{author}{\bibfnamefont{R.}~\bibnamefont{Bathe}},
  \bibinfo{author}{\bibfnamefont{S.~K.} \bibnamefont{Date}},
  \bibinfo{author}{\bibfnamefont{S.~I.} \bibnamefont{Patil}},
  \bibinfo{author}{\bibfnamefont{B.}~\bibnamefont{Hannoyer}},
  \bibinfo{author}{\bibfnamefont{F.}~\bibnamefont{Petit}}, \bibnamefont{and}
  \bibinfo{author}{\bibfnamefont{G.}~\bibnamefont{Marest}},
  \bibinfo{journal}{Phys.\ Rev.\ B} \textbf{\bibinfo{volume}{57}},
  \bibinfo{pages}{7841} (\bibinfo{year}{1998}).

\bibitem[{\citenamefont{Simopoulos et~al.}(1999)\citenamefont{Simopoulos,
  Pissas, Kallias, Devlin, Moutis, Panagiotopoulos, Niarchos, Christides, and
  Sonntag}}]{Simopoulos1999PRBa}
\bibinfo{author}{\bibfnamefont{A.}~\bibnamefont{Simopoulos}},
  \bibinfo{author}{\bibfnamefont{M.}~\bibnamefont{Pissas}},
  \bibinfo{author}{\bibfnamefont{G.}~\bibnamefont{Kallias}},
  \bibinfo{author}{\bibfnamefont{E.}~\bibnamefont{Devlin}},
  \bibinfo{author}{\bibfnamefont{N.}~\bibnamefont{Moutis}},
  \bibinfo{author}{\bibfnamefont{I.}~\bibnamefont{Panagiotopoulos}},
  \bibinfo{author}{\bibfnamefont{D.}~\bibnamefont{Niarchos}},
  \bibinfo{author}{\bibfnamefont{C.}~\bibnamefont{Christides}},
  \bibnamefont{and} \bibinfo{author}{\bibfnamefont{R.}~\bibnamefont{Sonntag}},
  \bibinfo{journal}{Phys.\ Rev.\ B} \textbf{\bibinfo{volume}{59}},
  \bibinfo{pages}{1263} (\bibinfo{year}{1999}).

\bibitem[{\citenamefont{Studer et~al.}(1999)\citenamefont{Studer, Toulemonde,
  Goedkoop, Barnabe, and Raveau}}]{Studer1999JJAPa}
\bibinfo{author}{\bibfnamefont{F.}~\bibnamefont{Studer}},
  \bibinfo{author}{\bibfnamefont{O.}~\bibnamefont{Toulemonde}},
  \bibinfo{author}{\bibfnamefont{J.}~\bibnamefont{Goedkoop}},
  \bibinfo{author}{\bibfnamefont{A.}~\bibnamefont{Barnabe}}, \bibnamefont{and}
  \bibinfo{author}{\bibfnamefont{B.}~\bibnamefont{Raveau}},
  \bibinfo{journal}{Jpn.\ J.\ Appl.\ Phys.\ Suppl.}
  \textbf{\bibinfo{volume}{38}}, \bibinfo{pages}{377} (\bibinfo{year}{1999}).

\bibitem[{\citenamefont{Kimura et~al.}(2001)\citenamefont{Kimura, Tokura,
  Kumai, Okimoto, and Tomioka}}]{Kimura2001JAP}
\bibinfo{author}{\bibfnamefont{T.}~\bibnamefont{Kimura}},
  \bibinfo{author}{\bibfnamefont{Y.}~\bibnamefont{Tokura}},
  \bibinfo{author}{\bibfnamefont{R.}~\bibnamefont{Kumai}},
  \bibinfo{author}{\bibfnamefont{Y.}~\bibnamefont{Okimoto}}, \bibnamefont{and}
  \bibinfo{author}{\bibfnamefont{Y.}~\bibnamefont{Tomioka}},
  \bibinfo{journal}{J.\ Appl.\ Phys.} \textbf{\bibinfo{volume}{89}},
  \bibinfo{pages}{6857} (\bibinfo{year}{2001}).

\end{thebibliography}
%
\end{document}